\newcommand{\be}{\begin{equation}}\newcommand{\ee}{\end{equation}}
\newcommand{\bea}{\begin{eqnarray}}\newcommand{\eea}{\end{eqnarray}}
\newcommand{\nn}{\nonumber}\newcommand{\p}[1]{(\ref{#1})}
 \newcommand{\lb}[1]{\label{#1}}

\newcommand\pab{\partial_{\alpha\beta}}

\newcommand\ab{{\alpha\beta}}

\newcommand\T{\mbox{Tr}}
\newcommand\s{\scriptscriptstyle}
\newcommand\A{\s A}
\newcommand\B{\s B}
\newcommand\C{\s C}
\newcommand\M{\s M}

\newcommand\R{\s R}
\newcommand{\pp}{\s ++}
\newcommand{\m}{\s --}
\newcommand{\0}{\s 0}
\newcommand{\2}{\s 2}
\newcommand{\3}{\s 3}
\newcommand{\4}{\s 4}

\newcommand{\Dp}{D^{\pp}}
\newcommand{\Dm}{D^{\m}}

\newcommand{\dpp}{\partial^{\pp}}
\newcommand{\dm}{\partial^{\m}}

\newcommand{\Doa}{D^{\0}_\alpha}

\newcommand{\Dpa}{D^{\pp}_\alpha}
\newcommand{\Dma}{D^{\m}_\alpha}

\newcommand{\DpS}{D^{\alpha\pp}D^{\pp}_\alpha}
\newcommand{\DmS}{D^{\alpha\m}D^{\m}_\alpha}
\newcommand{\DoS}{D^{\alpha\0}D^{\0}_\alpha}

\newcommand\dza{d\zeta^{\s(-4)}}
\newcommand{\Vp}{V^{\pp}}
\newcommand{\Vm}{V^{\m}}

\documentstyle[12pt]{article}

\begin{document}
\renewcommand{\thefootnote}{\fnsymbol{footnote}}
\hfill hep-th/9804167
\vspace{1cm}

\begin{center}
{\large\bf  HARMONIC SUPERSPACES FOR THREE-DIMENSIONAL  THEORIES
\footnote{This work is based on the talk given at the International 
Seminar "Supersymmetries and Quantum Symmetries"  dedicated to the 
memory of Prof. V.I. Ogievetsky (Dubna, July 22-26, 1997).}}
\vspace{0.5cm} \\
 B. M. Zupnik\footnote{On leave of absence from Institute of
Applied Physics, Tashkent State University, Uzbekistan.}
\vspace{0.2cm} \\
{\it Bogoliubov Laboratory of Theoretical Physics, JINR,  
 Dubna, Moscow Region, Russia}
 \vspace{1cm} \\
\end{center}
\begin{abstract}
Three-dimensional field theories with  $\;N=3\;$ and  $\;N=4\;$ 
supersymmetries are considered in  the framework of the harmonic-
superspace approach. Analytic superspaces of these supersymmetries are 
similar; however, the geometry of gauge theories with the manifest  
$\;N=3\;$ is richer and admits  construction of the topological mass term.
\end{abstract}

\renewcommand{\thefootnote}{\arabic{footnote}}
\setcounter{footnote}0
\setcounter{equation}0
\section{Introduction}
\hspace{0.5cm}
Three-dimensional supersymmetric gauge theories have been intensively 
studied in the framework of new nonperturbative methods in the field
theory \cite{IS,SW}. Superfield description of the simplest $\;D=3,~N=1,
2\;$ theories and various applications have earlier been  discussed in 
refs. [3 -- 10]. The most interesting features of $\;D=3\;$ theories are
connected with the Chern-Simons terms for  gauge fields and also with  
duality between vector and scalar fields.

We shall discuss the harmonic-superfield formalism of three-dimensional  
theories with the extended supersymmetries, which reflects the intrinsic
geometry  of these theories relevant for quantum description. The 
formulation of $\;D=3,~N=4\;$ superfield theories is analogous to the 
harmonic formalism of $\;D=4,~N=2\;$ theories \cite{GIK1,GI2,Z2}, although
the existence of different $SU(2)$ automorphism groups allows one to 
choose various versions of harmonic superspaces. General $N=3$ superfields
are not covariant with respect to  $N=4$ supersymmetry, but the analytic 
$N=3$ superfields are equivalent to the corresponding $\;N=4\;$ 
superfields. However, interactions with the manifest $\;N=3\;$ require a 
specific geometric description \cite{ZH,KL,Kao}, which does not guarantee 
conservation of the additional 4-th supersymmetry. The harmonic $\;N=3\;$
superspace of ref.\cite{ZH} has been based on the use of isovector 
harmonics, and now we consider the improved version of the harmonic 
formalism for these theories. Note that field models with $\;N=3\;$ 
supersymmetry (in distinction with $\;N=4\;$ models) are dual to the 
dynamics of non-orthogonal intersections of branes \cite{GGPT,OT}.

 Our conventions for the (2,1)-dimensional $\gamma$-matrices are
\be
(\gamma_m)_\ab (\gamma_n)^{\beta\rho}+
(\gamma_n)_\ab (\gamma_m)^{\beta\rho}=2\eta_{mn}\delta_\alpha^\rho~,\quad
(\gamma_m)_\ab \equiv \varepsilon_{\alpha\rho}(\gamma_m)_\beta^\rho~,
\lb{A1}
\ee
where $\eta_{mn}$ is the metric with signature $(1, -1, -1)$ and 
$\alpha, \beta\ldots$ are the $SL(2,R)$ spinor indices. We also shall use 
the basic notation of ref.\cite{GIK1} for the isospinor harmonics 
$u^\pm_i$.

\setcounter{equation}0
\section{N=4 harmonic superspace}
\hspace{0.5cm}
Superfield models with $\;D=3,~N=4\;$ supersymmetry can be studied via the
dimensional reduction of $\;D=4,~N=2\;$  superfield theories. We shall 
discuss $\;D=3,~N=4\;$ harmonic superspace by analogy with refs.
\cite{GIK1,GI2}. Let $z^{\M}=(x^{\alpha\beta}, \theta^\alpha_{ka})$ be
the {\it central} coordinates of the general $\;D=3,~N=4\;$ superspace 
$SS^4_3$, where $k, l\ldots$ and $a, b\ldots$ are  indices of the
automorphism groups $SU_{\s L}(2)$ and $SU_{\R}(2)$, respectively. In the
superspace without central charges, the  relations between basic spinor 
derivatives are
\be
\{ D^{ka}_\alpha, D^{lb}_\beta\}= 2i\varepsilon^{kl}\varepsilon^{ab}
\pab~,\lb{B1}
\ee
where $D^{ka}_\alpha=(D^k_\alpha~, \bar{D}^k_\alpha)$ and $\pab=
(\gamma^m)_\ab \partial_m$.

The superfield constraints of $\;N=4\;$ super-Yang-Mills theory $SYM^4_3$
can be written as follows:
\be
\{\nabla^{ka}_\alpha, \nabla^{lb}_\beta\}= 2i\varepsilon^{kl}
\varepsilon^{ab}\nabla_{\alpha\beta} + \varepsilon_{\ab}\varepsilon^{kl}
W^{ab}_{\s L}~, \lb{B2}
\ee
where $\nabla_{\M}$ are covariant derivatives with superfield connections
and $W^{ab}_{\s L}$ is a tensor superfield of the $SYM^4_3$ theory
 satisfying the constraint
\be
\nabla^{ia}_\alpha W^{bc}_{\s L} +\nabla^{ib}_\alpha W^{ca}_{\s L} +
\nabla^{ic}_\alpha W^{ab}_{\s L}=0~. \lb{B3}
\ee
Below we shall discuss  the alternative convention for the $SYM^4_3$
 constraints.

Introduce the notation  $W=W^{\s22}_{\s L}$ and $\overline{W}=
W^{\s11}_{\s L}$. The constraint for the Abelian gauge theory produces 
the following relations:
\bea
&&  D^k_\alpha \overline{W}=0~,\quad 
 \bar{D}^k_\alpha W=0~,\lb{B3b}\\
&& D^{k\alpha} D^l_\alpha W=\bar{D}^{k\alpha} \bar{D}^l_\alpha 
\overline{W}\lb{B3c}~,
\eea
which are analogous to the constraints of $\;N=2,~D=4\;$ vector multiplet.

One can also consider the on-shell constraints for different 
hypermultiplet superfields $q^{kb}_{\s L}$ and $q^{kb}_{\R}$
\bea
&&\nabla^{ia}_\alpha q^{kb}_{\s L} + \nabla^{ka}_\alpha q^{ib}_{\s L}=0~,
\lb{B4}\\
&&\nabla^{ia}_\alpha q^{kb}_{\R} + \nabla^{ib}_\alpha q^{ka}_{\R}=0~. 
\lb{B5}
\eea

The coset space of the automorphism group $SU(2)$ plays an important role
in the harmonic description of $\;N=2,~D=4\;$ superfield theory. In the 
harmonic approach to $\;N=4,~D=3\;$ theory we can use, alternatively, the
harmonic variables for cosets spanned on the generators  of the 
corresponding $SU(2)$ groups $L,~R$ or $(L+R)$. Let us  firstly consider
the harmonics $u^\pm_i$ for the group $SU_{\s L}(2)$. Using the standard 
harmonic methods of ref.\cite{GIK1} we can transform the  constraints 
(\ref{B2},\ref{B4}) into the following integrability ($L$-analyticity) 
condition:
\bea
&& \{\nabla^{+a}_\alpha,\nabla^{+b}_\beta\}=0~,\quad \nabla^{+a}_\alpha=
u^+_i\nabla^{ia}_\alpha~,\lb{B5b}\\
&&\nabla^{+a}_\alpha q^{+b}=0,~ \quad q^{+a}=u^+_iq^{ia}_{\s L}~. \lb{B6}
\eea

By analogy with the $\;D=4,~N=2\;$ case, one can use the $L$-analytic 
basis for the $SYM^4_3$ theory
\bea
&&\nabla^{+a}_\alpha=D^{+a}_\alpha=\partial/\partial \theta^{\alpha-}_a~,
\lb{B7b}\\
&&\nabla^{\pp}=\Dp + V^{\pp}_{\s L}~,\quad D^{+a}_\alpha V^{\pp}_{\s L}
=0~,\lb{B7}
\eea
where $D^{+a}_\alpha=(D^+_\alpha,~\bar{D}^+_\alpha) $ and $V^{\pp}_{\s L}$
is the  prepotential of $SYM^4_3$ depending on the  coordinates 
$\zeta_{\s L}=(x^{\ab}_{\s L},~\theta^{+\alpha}_a)$ of the $L$-analytic 
superspace $LSS^4_3$. In the physical gauge, it contains the components 
of the $\;D=3,~N=4\;$ vector multiplet
\bea
&& \Vp_{\s WZ}=(\theta^{+a}\theta^{+b}) \Phi_{(ab)}(x_{\s L})+
i(\theta^{+a}\gamma^m \theta^+_a) A_m(x_{\s L})+\nn\\
&& + (\theta^{+a}\theta^{+b})\theta^{+\alpha}_b u^-_k 
\lambda^k_{a\alpha}(x_{\s L})+ (\theta^+)^4 u^-_k u^-_l 
X^{(kl)}(x_{\s L})~.\lb{B8}
\eea

The solution of the constraint \p{B3} has the following form in the 
harmonic approach:
\be
W^{ab}_{\s L}=D^{+a\alpha}D^{+b}_\alpha \Vm_{\s L} (\Vp_{\s L})~,\lb{B8b}
\ee
where the standard solution for the 2-nd harmonic connection 
$\Vm_{\s L}$ \cite{Z2,Z4} is considered.

The analytic superfield $q^{+a}$ with the infinite number of auxiliary 
fields is the complete analog of the corresponding $\;D=4\;$ 
hypermultiplet representation. An alternative form of the analytic 
hypermultiplet can be obtained with the help of the harmonic duality 
transform $q^{+a}=u^{+a}\omega + u^{-a}F^{\pp} $ \cite{GI2}. On mass 
shell, these hypermultiplets have the following components:
\bea
&& q^{+a}_{\0}=u^+_k f^{ka}(x_{\s L})+ \theta^{+b\alpha}
\psi^a_{b\alpha}(x_{\s L})~,\lb{B6b}\\
&& \omega_{\0}=f(x_{\s L})+ u^+_k u^-_l f^{(kl)}(x_{\s L})+
\theta^{+b\alpha}u^-_k \psi^k_{b\alpha}(x_{\s L})~.\lb{B6c}
\eea

The holomorphic effective action of the Abelian $\;N=4\;$ gauge theory
contains the chiral superfield $W=\int du \;(\bar{D}^-)^2\Vp_{\s L}$ 
\cite{IZ}. One can consider the equivalent chiral and analytic 
representations of this action
\be
i\int d^{\3}x d^{\4}\theta\; {\cal F}(W) +\mbox{c.c.}=
i\int \dza_{\s L} du\; \Vp_{\s L}(D^+)^2\frac{{\cal F}(W)}{W}+
\mbox{c.c.}~, \lb{B8c}
\ee
where $\dza_{\s L}$ is the integral measure in $LSS^4_3$ and $d^{\4}
\theta$ is the spinor measure in the chiral superspace.

The  $u$-independent chiral superfield $A$ can be used for the 
construction of the complex analytic superfield $C^{\pp}=(D^+)^2 A$  
satisfying the additional harmonic constraint $\Dp C^{\pp}=0.$ 
The superfield $A$ is treated as a dual variable with respect to the 
Abelian 'magnetic' gauge superfield $\Vp_{\M}$ . The effective action of
this system contains $\Vp_{\M}$ as a Lagrange multiplier
\be
i\int d^{\3}x d^{\4}\theta\;{\cal F}(A) + i\int\dza du\;\Vp_{\M}(D^+)^2 A
+\mbox{c.c.}~. \lb{B8d}
\ee
One can obtain the reality constraint \p{B3c} for $A$ and the relation
\be
{\cal F}^\prime (A)=-\int du\; (\bar{D}^-)^2\Vp_{\M}\lb{B8e}
\ee
varying this action with respect to the superfields $\Vp_{\M}$ and $A$,
respectively. Note that the holomorphic representation with chiral 
superfields breaks the $SU_{\R}(2)$ automorphism group.

We shall not consider  here the classical action and quantization of the
$SYM^4_3$ theory since it can be obtained by a dimensional reduction from
the known $SYM^2_4$ theory.

Consider the Abelian case of the constraint \p{B5} for the 
$R$-hypermultiplet and define  the harmonic projection 
$R^{- a}\equiv u^{-}_k q^{ka}_{\R}$. In the $L$-analytic basis, the basic 
relations for this harmonic superfield  are
\be
 D^{+a}_\alpha R^{-b} + D^{+b}_\alpha R^{-a}=0~,\quad (\Dp)^2 R^{-a}=0~,
\lb{B9}
\ee
where the 1-st relation is treated as the constraint and the 2-nd one as
the equation of motion. Using the relation $\{ D^{+a}_\alpha, 
D^{-b}_\beta\}=-2i\varepsilon^{ab}\pab$ one can obtain a general covariant
solution of the constraint which contains the $L$-analytic bosonic and 
fermionic superfields $b^{-a}$ and  $f^\alpha$
\be
R^{-a}=b^{-a}+ D^{-a}_\alpha f^\alpha~. \lb{B10}
\ee
The harmonic equation of motion is equivalent to the equation 
$\Dm R^{-a}=0$. 

Thus, the hypermultiplet $R^{- a}$ (or its derivative $R^{+a}=\Dp R^{-a}$) 
is reduced to the pair of $L$-analytic superfields and their interactions 
can be described in $LSS^4_3$. On-shell it has the same components as the 
superfield $q^{ka}_{\R}$. 

One can identify the indices of the left and right automorphism groups
and use the $SU_{\C}(2)$-covariant spinor $\;N=4\;$ coordinates and 
supersymmetry parameters
\be
\theta^\alpha_{kl}=\theta^\alpha_{(kl)}+\varepsilon_{kl}
\;\theta^\alpha~,\quad
\epsilon^\alpha_{kl}=\epsilon^\alpha_{(kl)}+\varepsilon_{kl}
\;\epsilon^\alpha~,
\lb{B11}
\ee
where the isovector and isoscalar parts are introduced. The alternative 
$C$-form of the $SYM^4_3$ constraints contains the isovector
superfield $W^{kl}$
\be
\{\nabla^{km}_\alpha, \nabla^{ln}_\beta\}= 2i\varepsilon^{kl}
\varepsilon^{mn}\nabla_{\alpha\beta} + {1\over 2}\varepsilon_{\ab}
(\varepsilon^{kl}W^{mn} + 
\varepsilon^{mn}W^{kl})~. \lb{B12}
\ee

This representation allows us to separate the isoscalar covariant 
derivative 
\be
\{\nabla_\alpha, \nabla_\beta\}=i\nabla_{\alpha\beta}~,\quad
\{\nabla_\alpha, \nabla_\beta^{(kl)}\}=0~.\lb{B13}
\ee
The 2-nd relation is a conventional constraint which depends on a choice
of the $SU(2)$-frame.
Below we shall discuss  the commutation relations between the isovector 
$\;N=3\;$ covariant derivatives which are frame-independent.

\setcounter{equation}0
\section{New formulation of N=3 harmonic superspace}
\hspace{0.5cm}
Let us  consider now the new harmonic projections of the $~N=4~$ spinor 
coordinates \p{B11}
\be
\theta^{\alpha\s\pm\pm}=u^{\pm}_ku^{\pm}_l\theta^{\alpha kl}~,\quad
\theta^{\alpha\s\pm\mp}=u^{\pm}_ku^{\mp}_l\theta^{\alpha kl}~.
\lb{C0}
\ee
Coordinates of $LSS^4_3$ in the new 
representation are $\zeta_{\s L}=(x^{\ab}_{\s L},~\theta^{\alpha\pp},
~\theta^{\alpha\s+-})$ where 
$$
x^\ab_{\s L}=x^\ab +i(\theta^{\alpha\pp}\theta^{\beta\m}+\theta^{\beta\pp}
\theta^{\alpha\m}-\theta^{\alpha\s+-}\theta^{\beta\s-+}-
\theta^{\beta\s+-}\theta^{\alpha\s-+})~.
$$ 
 The infinitesimal $~N=4~$ spinor transformations have the following form
in these coordinates:
\bea
&& \delta x^{\ab}_{\s L}=\{2iu^{-k} u^{-l}\epsilon^\alpha_{(kl)}
\theta^{\beta\pp}+2i[\epsilon^\alpha-u^{-k} u^{+l}\epsilon^\alpha_{(kl)}]
\theta^{\beta\s+-}\}+\{\alpha\leftrightarrow \beta\}~,\lb{C0b}\\
&& \delta \theta^{\alpha\pp}=u^{+k} u^{+l}\epsilon^\alpha_{(kl)}~,\quad
\delta \theta^{\alpha\s+-}=\epsilon^\alpha + u^{+k} u^{-l}
\epsilon^\alpha_{(kl)}~,\lb{C0c}
\eea
where $\epsilon^\alpha$ is an isoscalar parameter of the 4-th 
supersymmetry.
Using the subgroup with $\epsilon^\alpha=0$ one can describe $N=3$
 supersymmetry in this $L$-analytic superspace.

A  three-dimensional $\;N=3\;$ supersymmetry can also be realized in the 
superspace $SS^3_3$ with the coordinates $z=(x^\ab, \theta^\alpha_{(kl)})
$. The corresponding superfields do not depend on the isoscalar 
coordinate $\theta^\alpha$ and are not covariant with respect to the 
4-th supersymmetry \footnote{In ref.\cite{ZH}, we have used the isovector 
$\;N=3\;$ spinor coordinates $\theta^\alpha_{\B}=(1/2)(\tau_{\B})^{kl}
\theta^\alpha_{(kl)}$.}. The constraints of $SYM^3_3$ in this superspace
are
\be
\{\nabla^{(km)}_\alpha, \nabla^{(ln)}_\beta\}= i(\varepsilon^{kl}
\varepsilon^{mn}+\varepsilon^{ml}\varepsilon^{kn})\nabla_{\alpha\beta} + 
{1\over 4}\varepsilon_{\ab}
(\varepsilon^{kl}W^{mn} +\varepsilon^{ml}W^{kn} + 
\varepsilon^{mn}W^{kl}+\varepsilon^{kn}W^{ml} )~, \lb{C0d)}
\ee
where all connections do not depend on $\theta^\alpha$.

Let us introduce the alternative analytic coordinates of the $\;N=3\;$ 
harmonic superspace $ASS^3_3$
\bea
&&x^\ab_{\A}=x^\ab +i(\theta^{\alpha\pp}\theta^{\beta\m}+
\theta^{\beta\pp}\theta^{\alpha\m})~,
\lb{C1}\\
&&\theta^{\alpha\pp}=u^{+}_ku^{+}_l\theta^{\alpha (kl)}~,\quad 
\theta^{\alpha\0}={1\over 2}(\theta^{\alpha\s+-}+\theta^{\alpha\s-+})=
u^{+}_ku^{-}_l\theta^{\alpha (kl)}~.\lb{C2}
\eea
It should be stressed that there is a one-to-one correspondence between 
the analytic $N=4$ and $N=3$ superfields. 

Spinor derivatives  have the following form in this $\;N=3\;$ superspace:
\bea
&& \Dpa=\partial^{\pp}_\alpha=\partial/\partial \theta^{\alpha\m}~,
\quad \Dma =\partial^{\m}_\alpha - 2i\theta^{\beta\m}\partial_\ab^{\A}~,
\lb{C3}\\
&& \Doa=-{1\over 2}\partial^{\0}_\alpha-i\theta^{\beta\0}
\partial_\ab^{\A}~,\quad \partial^{\s\pm\pm}_\alpha
\theta^{\beta\s\mp\mp}=\partial^{\0}_\alpha\theta^{\beta\0}=
\delta^\beta_\alpha~.\lb{C4}
\eea

The corresponding covariant harmonic derivatives are
\bea
&& \Dp=\dpp - 2i\theta^{\alpha\pp}\theta^{\beta\0}
\partial_\ab^{\A} + \theta^{\alpha\pp}\partial^{\0}_\alpha +
2\theta^{\alpha\0}\partial^{\pp}_\alpha~,\lb{C5}\\
&& \Dm=\dm + 2i\theta^{\alpha\m}\theta^{\beta\0}
\partial_\ab^{\A} + \theta^{\alpha\m}\partial^{\0}_\alpha +
2\theta^{\alpha\0}\partial^{\m}_\alpha~,\lb{C6}\\
&& [\Dm , \Dpa]=[\Dp , \Dma]=2\Doa~,\quad [D^{\s\pm\pm} , \Doa]= 
D^{\s\pm\pm}_\alpha~.\lb{C7}
\eea

Analytic $\;N=3\;$ superfields do not depend on $\theta^{\m}$ and are
unconstrained objects in $ASS^3_3$.

The $\;N=3\;$ covariant derivatives $\nabla^{(kl)}_\alpha$ in the central
basis can be transformed to the harmonized covariant  derivatives of the 
$SYM^3_3$ theory in the  basis with the analytic gauge group
\bea
&& \nabla^{\pp}_\alpha=\Dpa~,\quad \nabla^{\pp}=\Dp + \Vp~,\lb{C8}\\
&& \nabla^{\m} =\Dm + \Vm (\Vp)~,\quad \nabla^{\0}_\alpha=\Doa -
{1\over 2}\Dpa \Vm~, \lb{C9}\\
&& \nabla^{\m}_\alpha=[\nabla^{\m} , \nabla^{\0}_\alpha]~,\lb{C10}
\eea
where $\Vp$ is the analytic gauge prepotential in the adjoint 
representation of the gauge group, and $\Vm (\Vp)$ is the solution of the
 zero-curvature equation  for  harmonic connections \cite{Z2,Z4}.
In the physical $WZ$-gauge the prepotential contains the components
of the $\;N=3\;$ vector supermultiplet
\bea
&& \Vp_{\s WZ}=(\theta^{\pp})^2u^-_k u^-_l \Phi^{(kl)}(x_{\A})+
i(\theta^{\pp}\gamma^m \theta^{\0})A_m(x_{\A})+(\theta^{\0})^2
\theta^{\alpha\pp}\lambda_\alpha(x_{\A})+\nn\\
&& + (\theta^{\pp})^2\theta^{\alpha\0}u^-_k u^-_l
\lambda^{(kl)}_\alpha(x_{\A}) +(\theta^{\pp})^2(\theta^{\0})^2 u^-_k 
u^-_l X^{(kl)}(x_{\A})~,\lb{C10b}
\eea
which are analogous to the $\;N=4\;$ components \p{B8} with identified
$L$ and $R$ isospinor indices.

The basic superfield tensor of $SYM^3_3$ is analytic 
\be
W^{\pp}={1\over 2}\DpS\Vm
\lb{C11}
\ee
and satisfies the
additional $H$-constraint (Bianchi identity)
\be
\nabla^{\pp}W^{\pp}\equiv 0~.\lb{C11b}
\ee
It is a specific feature
of the $SYM^3_3$ theory that the prepotential $\Vp$ and its 
superfield-strength $W^{\pp}$ belong to the same analytic superspace 
$ASS^3_3$ .

Let us define  integral measures in the full and analytic $~N=3~$ harmonic
superspaces
\bea
&&d^9\!z_{\A}={1\over 64}d^3 x_{\A}(\DpS)(\DmS)(\DoS)~,\lb{C12b}\\
&&\dza ={1\over 16}d^3 x_{\A} (\DmS)(\DoS)~.
\lb{C12}
\eea
Note that these measures have dimensions $~d=0~$ and $~1$ ,
respectively.

The standard kinetic term of the $SYM^3_3$ action is
\be
S_k={1\over g^2}\int \dza du\; \T\; W^{\pp}W^{\pp}~, \lb{C13}
\ee
where $g$ is the coupling constant with dimension $d=-1/2$.

The  effective action of the Abelian $~N=3~$ theory contains
an arbitrary function of the $H$-constrained superfield $W^{\pp}$ 
\be
\int \dza du \;{\cal G}^{\s(+4)}(W^{\pp},u)=\int \dza du\;[\tau 
(W^{\pp})^2
+ \sum\limits_{p=1}^{\infty}c^{l_1...l_{2p}}u^-_{l_1}\ldots u^-_{l_{2p}} 
(W^{\pp})^{p+2} ]~, \lb{C14}
\ee
where $\tau, c^{l_1...l_{2p}}$ are some constants. It is clear that only
quadratic term of the general action conserves the $SU_{\C}(2)$ symmetry.
This analytic representation of the low-energy effective action is 
alternative to the holomorphic $N=4$ representation \p{B8c}.

The interaction of the gauge superfield $W^{\pp}(\Vp)$ is dual to the 
following interaction of the unconstrained real analytic superfields 
$\omega$ and $A^{\pp}$:
\be
\int \dza du\; [{\cal G}^{\s(+4)}(A^{\pp},u) + A^{\pp}\Dp \omega]~.
\lb{C15}
\ee
Varying $\omega$ yields the constraint $\Dp A^{\pp}=0$. This action is 
the first-order form of the special interaction of $\omega$ and 
$\Dp\omega$, although the elimination of the superfield $A^{\pp}$ is a 
non-trivial algebraic problem for the general function 
${\cal G}^{\s(+4)}$.

An important feature of the $SYM^3_3$ theory is the existence of a 
topological mass (Chern-Simons) term \cite{ZH}. In the improved $~N=3~$
harmonic formalism this term can be constructed by the analogy with the 
action of $SYM^2_4$ \cite{Z4}
\be
 S_m ={m\over g^2}\sum\limits^{\infty}_{n=2}\frac{(-1)^{n+1}}{n} 
\int d^9\!z du_{\s1} \ldots du_n
\frac{\T\; [\Vp(z,u_{\s1})\ldots \Vp(z,u_n)]}{(u_{\s1}^+ u_{\2}^+)
\ldots (u_n^+u_{\s1}^+)} \label{C16}
 \ee
where $(u_{\s1}^+ u_{\2}^+)^{-1}$ is the harmonic distribution \cite{GI2}.
 Note that the measure $d^9\!z$ in this term is not covariant with 
respect to the 4-th supersymmetry. The analytic version of the topological 
mass term and the  Fayet-Iliopoulos term for the Abelian theory has the 
following form:
\be
S_m+S_{\s FI}={1\over g^2}\int \dza du\;(m\Vp W^{\pp}+\xi^{\pp}\Vp)~. 
\lb{C16b}
\ee 

The action $S_k+S_m+S_{\s FI}$ yields the free equation of motion for the 
$~N=3~$ Abelian gauge theory 
\be
[(D^{\alpha\0}\Doa)+m]\;W^{\pp}+\xi^{\pp}=0~.
 \lb{C18}
\ee
This equation has the following vacuum solutions:
\bea
 W^{\pp}=-{1\over m}\xi^{\pp}\quad \mbox{for}\;m\neq 0~,
\hspace{3cm}&&\lb{C19}\\
 W^{\pp}=a^{(kl)}u^+_k u^+_l-2[(\theta^{\0})^2\xi^{\pp}-
2(\theta^{\pp}\theta^{\0})\xi^{\0}+(\theta^{\pp})^2\xi^{\m}]\quad
 \mbox{for}\;m=0~,&&
\lb{C20}
\eea
where $\xi^{\s\pm\pm}=\xi^{(kl)}u^{\pm}_k u^{\pm}_l~,~\xi^{\0}=
(1/2)\Dp\xi^{\m}$ and $a^{(kl)}$ and $\xi^{(kl)}$ are arbitrary constants. 
Note that the first solution does not break supersymmetry.

For the case $m=0,~\xi^{(kl)}=0$ we can study  
the background Abelian prepotential
\be
V^{\pp}={1\over 2}a^{(kl)}[(\theta^{\0})^2 u^+_k u^+_l-
2(\theta^{\pp}\theta^{\0})u^+_k u^-_l+(\theta^{\pp})^2u^-_k u^-_l]~,
\lb{C21}
\ee
which introduces the $N=3$  central charges and produces masses of 
charged superfields by analogy with ref.\cite{IKZ}.

The minimal gauge interaction of the $q^+$ hypermultiplet has the
 following form:
\be
\int \dza du\; \bar{q}^+(\Dp + \Vp)q^+~.\lb{C17}
\ee
The free hypermultiplet satisfies the equation $\Dp q^+ =0$ and contains 
a finite number of complex on-shell components 
\be
 q^+_{\0}=u^+_k f^k(x_{\A})+ (\theta^{\alpha\pp}u^-_k-
\theta^{\alpha\0}u^+_k)\psi^k_\alpha(x_{\A})~.\lb{C17b}
\ee
The real $~N=3~~\omega$-hypermultiplet has been described in ref.
\cite{ZH}.

Note that the similar harmonic methods can be used for description of
two-dimensional models with (3,3) supersymmetry.

The author is grateful to E. Ivanov and N. Ohta for stimulating 
discussions.
This work is partially supported  by  grants  RFBR-96-02-17634, 
RFBR-DFG-96-02-00180,  INTAS-93-127-ext and INTAS-96-0308, and
by  grant of Uzbek Foundation of Basic Research N 11/97.

\end{document}